\documentclass[conference]{IEEEtran}

\IEEEoverridecommandlockouts

\usepackage[T1]{fontenc}
\usepackage[latin1]{inputenc}
\usepackage[english]{babel}
\usepackage{booktabs,verbatim,multirow,etoolbox}
\usepackage{color}
\usepackage{cite}
\usepackage{amsmath,amssymb,amsfonts}
\usepackage{algorithmic}
\usepackage{textcomp}
\usepackage{balance}
\usepackage{xspace}
\usepackage[pdftex]{graphicx}
\usepackage{orcidlink}

\usepackage{fixltx2e}

\begin{document} 

\title{Geometrically-Motivated Primary-Ambient Decomposition With Center-Channel Extraction}
\author{\IEEEauthorblockN{Jouni Paulus\IEEEauthorrefmark{1}~\orcidlink{0000-0003-2283-2062}\thanks{\IEEEauthorrefmark{1}Now with WS Audiology, Erlangen, Germany.}}
\IEEEauthorblockA{\textit{Fraunhofer Institute for Integrated Circuits IIS}, and \\
\textit{International Audio Laboratories Erlangen\IEEEauthorrefmark{2}}\thanks{\IEEEauthorrefmark{2}International Audio Laboratories Erlangen is a joint institution of Universit{\"a}t Erlangen-N{\"u}rnberg and Fraunhofer IIS.}\\
Erlangen, Germany}
\and
\IEEEauthorblockN{Matteo Torcoli~\orcidlink{0000-0003-2834-9194}}
\IEEEauthorblockA{\textit{Fraunhofer Institute for Integrated Circuits IIS}, and \\
\textit{International Audio Laboratories Erlangen\IEEEauthorrefmark{2}} \\
Erlangen, Germany}}

\newcommand{\mathbi}[1]{\textbf{\em #1}}

\newcommand{\sigL}{\ensuremath{\mathbi{l}}}
\newcommand{\sigR}{\ensuremath{\mathbi{r}}}
\newcommand{\sigC}{\ensuremath{\mathbi{c}}}

\newcommand{\obsL}{\ensuremath{\mathbi{x}_L}}
\newcommand{\obsR}{\ensuremath{\mathbi{x}_R}}

\newcommand{\ambL}{\ensuremath{\mathbi{a}_L}}
\newcommand{\ambR}{\ensuremath{\mathbi{a}_R}}
\newcommand{\primaryL}{\ensuremath{\mathbi{p}_L}}
\newcommand{\primaryR}{\ensuremath{\mathbi{p}_R}}

\newcommand{\remixLf}{\ensuremath{\mathbi{y}_{L,front}}}
\newcommand{\remixRf}{\ensuremath{\mathbi{y}_{R,front}}}
\newcommand{\remixLr}{\ensuremath{\mathbi{y}_{L,rear}}}
\newcommand{\remixRr}{\ensuremath{\mathbi{y}_{R,rear}}}

\newcommand{\obsC}{\ensuremath{\mathbi{C}_x}}
\newcommand{\crossC}{\ensuremath{\mathbi{C}_{s,x}}}
\newcommand{\rotC}{\ensuremath{\mathbi{C}_{rot}}}
\newcommand{\covEl}[1]{\ensuremath{c_{#1}}}
\newcommand{\objCovMat}{\ensuremath{\mathbi{C}_{obj}}}

\newcommand{\dmxMat}{\ensuremath{\mathbi{D}}}
\newcommand{\unmixMat}{\ensuremath{\mathbi{G}}}
\newcommand{\rotMat}{\ensuremath{\mathbi{R}_{\theta}}}

\newcommand{\oneObj}{\ensuremath{\mathbi{s}}}
\newcommand{\objMat}{\ensuremath{\mathbi{S}}}
\newcommand{\estObjMat}{\ensuremath{\hat{\mathbi{S}}}}
\newcommand{\oneMixSig}{\ensuremath{\mathbi{y}}}
\newcommand{\mixSigMat}{\ensuremath{\mathbi{Y}}}
\newcommand{\unmixRow}{\ensuremath{\mathbi{g}}}
\newcommand{\costFun}{\ensuremath{J}}

\maketitle

\begin{abstract}
A geometrically-motivated method for primary-ambient decomposition is proposed and evaluated in an up-mixing application. 
The method consists of two steps, accommodating a particularly intuitive explanation.
The first step consists of signal-adaptive rotations applied on the input stereo scene, which translate the primary sound sources into the center of the rotated scene. 
The second step applies a center-channel extraction method, based on a simple signal model and optimal in the mean-squared-error sense.
The performance is evaluated by using the estimated ambient component to enable surround sound starting from real-world stereo signals. 
The participants in the reported listening test are asked to adjust the audio scene envelopment and find the audio settings that pleases them the most. 
The possibility for up-mixing enabled by the proposed method is used extensively, and the user satisfaction is significantly increased compared to the original stereo mix.
\end{abstract}


\section{Introduction}
\label{sec:intro}
Center-channel extraction (CE) and primary-ambient decomposition (PAD) are audio signal processing algorithms commonly used for decomposing a 2-channel input signal into a 3- or 4-channel output.
The output can consist of estimates of the left, right, and center channel, or primary (or direct) components with a direction and ambient components without a clear direction. 
The methods are often motivated by the applications including dereverberation or controlling dry/wet-ratio~\cite{Allen77-JASA, Walther11-WASPAA, Uhle15-ICASSP, Goodwin08-Asilomar}, signal separation~\cite{Geiger15-EUSIPCO, Barry04-DAFX, Paulus19-JAES}, up-mixing from stereo to a higher number of output channels~\cite{Usher07-TALSP, Faller06-JAES, Baek12-AES, Harma11-EUSIPCO, Harma11-AES, Avendano02-ICASSP, Irwan02-JAES, Ibrahim18-ICASSP, Kraft16-AES, Hashimoto18-AES}, direct source re-panning without modifying the ambience~\cite{Adami15-EUSIPCO}, and audio coding by separating spatial cues from the audio itself~\cite{Goodwin07-ICASSP}.

With the exception of some more recent deep-learning based approaches, e.g.,~\cite{Ibrahim18-ICASSP, Lim20-BMSB}, the CE and PAD solutions rely on the cross-correlation properties of the observed stereo signal channels and the signal model assumptions. 
One of the simplest forms is from~\cite{Avendano02-ICASSP}, where the cross-channel coherence value of the input is mapped into an ambience separation mask weight.
Other methods attempt to estimate the covariance properties of the target component signal with explicit equations, e.g.,~\cite{Allen77-JASA, Jourjine00-ICASSP, Uhle15-ICASSP, Walther11-WASPAA, Harma11-AES, Vickers09-AES, Geiger15-EUSIPCO, Faller06-JAES}, or by utilizing principal component analysis (PCA).
The PCA methods, e.g.,~\cite{Irwan02-JAES} assume that the eigenvalues of the observation covariance matrix correspond to the levels of the primary and ambient components~\cite{Goodwin07-ICASSP}.
Assuming the larger eigenvalue to correspond to the primary component may lead to under-estimating the ambient component level and some works use scaling based on the ratio of the eigenvalues~\cite{Goodwin08-ICASSP, Goodwin08-Asilomar, Baek12-AES, Ibrahim16-SMC}.
The separation may rely on per-channel filters, e.g.,~\cite{Merimaa07-AES}, common filters for all channels, e.g.,~\cite{Walther11-WASPAA}, or cross-channel filters, e.g.,~\cite{Usher07-TALSP}.
Targeting at signals with dominant ambient components,~\cite{He14-SPL} formulates the PAD problem into a task of finding ambient component phase maximizing the sparsity of the primary component.
This model was recently extended by introducing an assumption of the primary component having a harmonic spectrum~\cite{Zhu21-EUSIPCO}.
Some works extend the definition from two input channels into multiple channels and defining the direct sound to be the part common in all channels~\cite{Uhle14-AES}.
Some methods have multiple methods running in parallel and use late fusion to outperform the single approaches~\cite{He16-AES}.
The work~\cite{He14-TASLPACM} makes a good effort of unifying several approaches a common vocabulary and analysing the methods more closely. 

The contribution of this paper is threefold.
1) We present a step-by-step derivation of a multi-channel Wiener-filter CE method based on estimating the signal statistics from a simple signal model.
2) This method is extended for a novel PAD solution by adding geometrically-motivated pre- and post-processing rotation steps.
3) The performance of the proposed PAD method is evaluated with the Adjustment/Satisfaction Test (A/ST) in a stereo-to-quad up-mixing application.

\section{Background}
\label{sec:background}
The CE methods attempt to decompose a 2-channel (stereo) signal $\obsL$ and $\obsR$ into three components: left $\sigL$, center $\sigC$, and right $\sigR$ based on the signal model (the derivation assumes these being possibly complex-valued sub-band signals, e.g., in short-time Fourier transform (STFT) domain), and the time-dependency and band index are dropped in the following):
\begin{equation}
\label{eq:observation}
\begin{cases}
\obsL = \sigL + \sigC\\
\obsR = \sigR + \sigC
\end{cases},
\end{equation}
which in a matrix form is:
\begin{equation}
\label{eq:mixModel}
\begin{bmatrix}
\obsL^\top \\
\obsR^\top 
\end{bmatrix}
=
\dmxMat
\begin{bmatrix}
\sigL^\top \\
\sigR^\top \\
\sigC^\top
\end{bmatrix},
\end{equation}
where $\mathbi{x}^\top$  is the transpose of $\mathbi{x}$ and $\dmxMat$ is the mixing matrix:
\begin{equation}
\dmxMat = 
\begin{bmatrix}
1 & 0 & 1 \\
0 & 1 & 1 
\end{bmatrix}.
\end{equation}
An assumption making the decomposition more approachable is that all the three component signals are statistically independent, i.e., 
\begin{equation}
\label{eq:components}
E\{\sigL^H \sigR\} = E\{\sigL^H \sigC\} = E\{\sigC^H \sigR\} = 0,
\end{equation}
where $E\{\cdot\}$ is the expected value operator over time and where $\mathbi{x}^H$  is the Hermitian transpose

The PAD methods use the signal model:
\begin{equation}
\label{eq:PADobservation}
\begin{cases}
\obsL = \ambL + \primaryL\\
\obsR = \ambR + \primaryR
\end{cases},
\end{equation}
where $\ambL$ and $\ambR$ are the ambient components, and $\primaryL$ and $\primaryR = g\primaryL$ the primary signal components, all to be estimated by the method, and $g$ is a level-panning weight.
Similar to~\eqref{eq:components}, the component signals are assumed to be independent, i.e., 
\begin{equation}
E\{\ambL^H \ambR\} = E\{\ambL^H \primaryL\} = E\{\ambR^H \primaryR\} = 0.
\end{equation}
Some earlier works, e.g.,~\cite{Harma11-EUSIPCO, Uhle15-ICASSP} on PAD make the assumption that the ambient signal components have the same level, i.e., $E\{\ambL^H \ambL\} = E\{\ambR^H \ambR\}$.
The PAD algorithm proposed in this paper relaxes this assumption.

\section{Proposed Method}
\label{sec:method}
This section describes the estimation of the signal statistics from the observations, a minimum-mean-squared error (MMSE) method using the signal statistics for performing CE, and finally a novel geometry-based extension of the CE-solution for PAD.

\subsection{Estimating Signal Statistics}
\label{sec:statistics}
The proposed CE-method needs to know the covariance matrix $\objCovMat$ of the component signals. 
Remembering the assumption that the three component signals are independent~\eqref{eq:components}, the covariance matrix consists of only the component energies $E\{\sigL^H \sigL\}$, $E\{\sigR^H \sigR\}$, and $E\{\sigC^H \sigC\}$ as 
\begin{equation}
\label{eq:compEnergies}
\objCovMat = 
\begin{bmatrix}
E\{\sigL^H \sigL\} & 0 & 0 \\
0 & E\{\sigR^H \sigR\} & 0 \\
0 & 0 & E\{\sigC^H \sigC\}
\end{bmatrix}.
\end{equation}

The (auto- and cross-) covariance matrix of the two observed mixture signals is
\begin{equation}
\label{eq:obsCovariance}
\obsC = 
\begin{bmatrix}
\covEl{L,L} & \covEl{L,R}\\
\covEl{R,L} & \covEl{R,R}
\end{bmatrix}
=
\begin{bmatrix}
E\{\obsL^H \obsL\} & E\{\obsL^H \obsR\} \\
E\{\obsR^H \obsL\} & E\{\obsR^H \obsR\}
\end{bmatrix}.
\end{equation}
Here, $\covEl{i,j} = E\{\mathbi{x}_i^H \mathbi{x}_j\}$ are the covariance matrix entries between the two observed signals $\mathbi{x}_i$ and $\mathbi{x}_j$, when we assume the two signals to have zero mean.
With the assumption on the statistical independence of the component signals, the covariance matrix mixing model is
\begin{equation}
\begin{bmatrix}
\covEl{L,L} \\
\covEl{L,R} \\
\covEl{R,L} \\
\covEl{R,R}
\end{bmatrix} = 
\begin{bmatrix}
1 & 1 & 0 \\
0 & 1 & 0 \\
0 & 1 & 0 \\
0 & 1 & 1 
\end{bmatrix} 
\begin{bmatrix}
E\{\sigL^H \sigL\} \\
E\{\sigC^H \sigC\} \\
E\{\sigR^H \sigR\} 
\end{bmatrix}.
\end{equation}
Solving this provides us the estimates of the three component signal energies as
\begin{equation}
\begin{cases}
E\{\sigL^H \sigL\} = \covEl{L,L} - \mathcal{R}\{\covEl{L,R}\} \\
E\{\sigC^H \sigC\} = \mathcal{R}\{\covEl{L,R}\} \\
E\{\sigR^H \sigR\} = \covEl{R,R} - \mathcal{R}\{\covEl{L,R}\}
\end{cases},
\end{equation}
where $\mathcal{R}\{x\}$ returns the real part of the complex number. 
For the presentation clarity, the operator is omitted from the equations in the remainder of this paper.

\subsection{Component Separation}
\label{sec:mmse}
The component signals can be obtained with a multi-channel Wiener filter $\unmixMat$ applied on the observed stereo signal as 
\begin{equation}
\begin{bmatrix}
\hat{\sigL}^\top \\
\hat{\sigR}^\top \\
\hat{\sigC}^\top
\end{bmatrix}
= \unmixMat 
\begin{bmatrix}
\obsL^\top \\
\obsR^\top
\end{bmatrix}.
\end{equation}
The filter coefficients can be obtained using a well-known MMSE solution
\begin{equation}
\label{eq:unmixMat}
\unmixMat = \crossC \obsC^{-1},
\end{equation}
where $\crossC$ is a covariance matrix between the components and mixture signals.
From~\eqref{eq:mixModel} this can be written as the product
\begin{equation}
\crossC = \objCovMat \dmxMat^H .
\end{equation}
Taking the component signal covariance matrix elements from~\eqref{eq:compEnergies}, this can be written as
\begin{equation}
\crossC = 
\begin{bmatrix}
\covEl{L,L} - \covEl{L,R} & 0 \\
0 & \covEl{R,R} - \covEl{L,R} \\
\covEl{L,R} & \covEl{L,R}
\end{bmatrix}.
\end{equation}
This leads into the full un-mixing matrix solution of
\begin{equation}
\unmixMat = \frac{\begin{bmatrix}
	\covEl{L,L}\covEl{R,R} - \covEl{R,R}\covEl{L,R} & \covEl{L,R}^2 - \covEl{L,R}\covEl{L,L} \\
	\covEl{L,R}^2-\covEl{L,R}\covEl{R,R}            & \covEl{L,L}\covEl{R,R} -\covEl{L,L}\covEl{L,R} \\
	\covEl{R,R}\covEl{L,R} -\covEl{L,R}^2           & \covEl{L,L}\covEl{L,R} - \covEl{L,R}^2
	\end{bmatrix}}{\covEl{L,L}\covEl{R,R} - \covEl{L,R}^2}.
\end{equation}

\subsection{Extension into Primary-Ambient Decomposition}
\label{sec:PAD}
For applying the CE algorithm derived above for solving the PAD model of~\eqref{eq:PADobservation} we assume that our observation is a spatially rotated CE-scene in which the dominant sound direction corresponds to the primary / direct component.
In other words, we find a rotation angle $\theta$ such that when applied on our 2-channel observation $\obsL$, $\obsR$, the resulting rotated scene fulfils the assumptions that a) the primary signal component is centered in the scene and b) uncorrelated with the two independent ambient signal components c) that have equal energies.
We can apply CE in this rotated domain, and then apply inverse rotation on the extracted components.

The rotation of the scene is done with a rotation matrix
\begin{equation}
\label{eq:rotMat}
\rotMat = 
\begin{bmatrix}
\cos\theta& -\sin\theta \\
\sin\theta  & \cos\theta
\end{bmatrix}.
\end{equation}
The rotation angle $\theta$ can be solved from the covariance matrix of the rotated observations, being equal to applying the rotation on the covariance matrix of the observations:
\begin{equation}
\label{eq:rotCov}
\rotC = 
\rotMat \obsC \rotMat^H
=
\begin{bmatrix}
\covEl{1,1} & \covEl{1,2} \\
\covEl{2,1} & \covEl{2,2}
\end{bmatrix}.
\end{equation}
Requiring that the rotated observed signals have the same energy corresponds to requiring the main diagonal elements to be equal: $\covEl{1,1} = \covEl{2,2}$.
Combining~\eqref{eq:obsCovariance} and~\eqref{eq:rotCov} and solving the rotation from this gives us the angle
\begin{equation}
\label{eq:theta}
\theta = \frac{1}{2} \tan^{-1} \frac{\covEl{L,L} - \covEl{R,R,}}{2\covEl{L,R}}.
\end{equation}

Now we can use the earlier center-channel extraction algorithm for the decomposition.
The full proposed PAD algorithm consists of the steps:
\begin{enumerate}
\item Compute input covariance matrix $\obsC$ with~\eqref{eq:obsCovariance}.
\item Determine the centering rotation $\theta$ with~\eqref{eq:theta}, and compute the rotated covariance matrix $\rotC$ with~\eqref{eq:rotCov}.
\item Compute the un-mixing matrix $\unmixMat$ with~\eqref{eq:unmixMat} using the rotated covariance matrix $\rotC$ in the place of the observation covariance matrix $\obsC$.
\item Combine the centering rotation $\rotMat$ and the counter-rotation $\rotMat^H$ of the separated components (single-channel primary component duplicated in both outputs) with the un-mixing matrix $\unmixMat$ using
\begin{equation}
\label{eq:unmix:full}
\unmixMat_{full} = 
\begin{bmatrix}
\rotMat^H & \mathbi{0} \\
\mathbi{0} & \rotMat^H
\end{bmatrix}
\begin{bmatrix}
1 & 0 & 0 \\
0 & 1 & 0 \\
0 & 0 & 1 \\
0 & 0 & 1
\end{bmatrix}
\unmixMat \rotMat.
\end{equation}
(The mixing matrix in the equation assigns the center of the rotated scene into the primary signal.)
\item Apply the combined un-mixing matrix on the observed signals with
\begin{equation}
\begin{bmatrix}
\ambL^\top \\
\ambR^\top \\
\primaryL^\top \\
\primaryR^\top
\end{bmatrix}
= \unmixMat_{full}
\begin{bmatrix}
\obsL^\top \\
\obsR^\top
\end{bmatrix}.
\end{equation}
\end{enumerate}

The counter-rotation achieves that the primary component is located again in the original direction.
A possible drawback of the counter-rotation applied on the ambient components is that even if the extracted components would be independent and of equal energy in the rotated coordinate system, the counter-rotation modifies both these properties.

It is worth noting that the computation of the un-mixing matrix~\eqref{eq:unmix:full} does not need to apply the rotations to the observed signals, but all operations can be performed in the covariance matrix domain.
Further computational complexity optimizations can be obtained by inspecting the structure of the full un-mixing matrix more closely.
We divide it into two parts corresponding to the ambient and primary components:
\begin{equation}
\unmixMat_{full} = 
\begin{bmatrix}
\unmixMat_A \\
\unmixMat_P
\end{bmatrix}.
\end{equation}
The part corresponding to the ambient signal component is 
\begin{align}
\unmixMat_A &= \rotMat^H
\begin{bmatrix}
1 & 0 & 0 \\
0 & 1 & 0
\end{bmatrix}
\unmixMat \rotMat \\
&=
\begin{bmatrix}
\covEl{R,R} & -\covEl{L,R} \\
-\covEl{L,R} & \covEl{L,L}
\end{bmatrix}
\frac{k - \covEl{L,L} - \covEl{R,R}}{2 (\covEl{L,R}^2 - \covEl{L,L} \covEl{R,R})},
\end{align}
with
\begin{equation}
k = \sqrt{(\covEl{L,L} - \covEl{R,R})^2 + 4\covEl{L,R}^2}.
\end{equation}
In earlier publications, e.g.,~\cite{Walther11-WASPAA, Faller06-JAES, Merimaa07-AES} this quantity corresponds to the total energy of the direct component in their signal model.

The part corresponding to the primary signal component is
\begin{equation}
\unmixMat_P = \rotMat^H
\begin{bmatrix}
0 & 0 & 1 \\
0 & 0 & 1
\end{bmatrix}
\unmixMat \rotMat \\
=
\begin{bmatrix}
g_{P1,1} & g_{P1,2} \\
g_{P2,1} & g_{P2,2}
\end{bmatrix}
\end{equation}
with
\begin{equation}
\begin{cases}
g_{P1,1} = \frac{2\covEl{L,R}^2 + \covEl{R,R}(\covEl{R,R} - \covEl{L,L} - k) }{2 (\covEl{L,R}^2 - \covEl{L,L} \covEl{R,R})} \\
g_{P1,2} = \frac{-\covEl{L,R} (k - \covEl{L,L} - \covEl{R,R})}{2 (\covEl{L,R}^2 - \covEl{L,L} \covEl{R,R})} \\
g_{P2,1}  = g_{P1,2} \\
g_{P2,2} = \frac{2\covEl{L,R}^2 + \covEl{L,L}(\covEl{L,L} - \covEl{R,R} - k) }{2 (\covEl{L,R}^2 - \covEl{L,L} \covEl{R,R})}
\end{cases}.
\end{equation}
It is possible to manipulate this representation further and to find the relationship
\begin{equation}
\unmixMat_P = \mathbi{I} - \unmixMat_A,
\end{equation}
where $\mathbi{I}$ is a 2-by-2 identity matrix.

Considering the covariance matrix of the rotated observed signals~\eqref{eq:rotCov}, and inserting the computed rotation angle from~\eqref{eq:theta}, the cross-terms can be simplified into 
\begin{equation}
\covEl{1,2} = \covEl{2,1} = \frac{1}{2}\sqrt{(\covEl{L,L}-\covEl{R,R})^2 + 4\covEl{L,R}^2} = \frac{1}{2} k .
\end{equation}

\section{Evaluation}
\label{sec:eval}
The proposed method is evaluated in a listening test focusing on the up-mixing application, similar to~\cite{Usher07-TALSP, Ibrahim18-ICASSP}.
The stereo input is split into primary and ambient components and some of the ambient component can be moved into the rear channels of a 5.1 setup (the center and LFE channels stay silent). 
The amount moved is controlled by the user with a dial over 15 steps from full signal in the front channels to only primary component in the front and ambient component in the rear channels.
The range is extended by 10 steps in which the ambient component is further amplified up to 20\,dB, and in the low end by 5 steps of gradual stereo image narrowing to dual mono, see Table~\ref{table:upmix}.
All conditions are normalized to equal integrated loudness~\cite{ITU15-BS1770}.
The adjustment is done with a continuously rotating dial without a tactile feedback of the steps. 
The test is conducted using the Adjustment Satisfaction Test (A/ST)~\cite{Torcoli18-TBC, Paulus19-JAES} with the main task of \emph{"Please find the audio setting that pleases you the most with respect to audio scene envelopment and overall audio quality."}
After adjusting the audio setting, the listener rates their satisfaction of the adjusted version compared to the original stereo mix on the scale \emph{much worse}, \emph{worse}, \emph{slightly worse}, \emph{same}, \emph{slightly better}, \emph{better}, \emph{much better}, each with 5 steps between two labelled positions on the scale.

\begin{table}[t]
\caption{The loudspeaker signals in the three operating regions in the up-mixing listening test.
In the \emph{narrowing} range the stereo image is made narrower with the cross-mixing coefficient $a \in$\{0.5, 0.57, 0.66, 0.76, 0.87\}.
In the \emph{amb. relocation} region the ambient component in the front channels is attenuated by a gain $g_a \in$ \{0, -1.5, -3, -5, -7.5, -10.5, -14, -18, -23, -28, -34, -41, -49, -59, -76, -96\}\,dB, and the matching contribution is moved to the rear channels.
In the \emph{amb. boost} region, the full ambient signal is located in the rear channels and further amplified by \mbox{$b_a \in$ \{1, 3, 5, 7, 9, 11, 13, 15, 17, 20\}}\,dB.
}
\vspace{-0.5em}
\label{table:upmix}
\centering
\begin{tabular}{c|c|c|c}
  output channel & \textbf{narrowing} & \textbf{amb. relocation} & \textbf{amb. boost} \\ 
\hline 
$\remixLf$ & $a \obsL + (1-a) \obsR$  & $\primaryL + g_a \ambL$ & $\primaryL$ \\ 
$\remixRf$ & $(1-a) \obsL + a \obsR$  & $\primaryR + g_a \ambR$ & $\primaryR$ \\ 
$\remixLr$ & $0$                      & $(1 - g_a) \ambL$       & $b_a \ambL$ \\ 
$\remixRr$ & $0$                      & $(1 - g_a) \ambR$       & $b_a \ambR$ \\ 
\end{tabular}
\vspace{-1em}
\end{table}


The test stimuli are 16 items of 20--30\,s in length, half are music-only (instrumental only, e.g., orchestral, and with singing, e.g., pop) and half are real-world broadcast content (TV dramas and magazine shows, where speech overlaps with background music, effects, or noise).
The sampling rate is 48\,kHz.
The processing uses STFT with 1024-sample frames with 50\% overlap, sine window, and 2x zero-padding. 
The observation cross-covariance matrix is computed for each frequency bin per-frame, and these are averaged with a 5-frame sliding mean for obtaining~\eqref{eq:obsCovariance}.
The computed un-mixing matrices $\unmixMat_P$ and $\unmixMat_A$ are smoothed with a 3-frame sliding mean for reducing musical noise.
The test takes place in a listening room with a 5.1 loudspeaker setup.

We compared informally several PAD algorithms from the literature, e.g.,~\cite{Uhle15-ICASSP, Ibrahim16-SMC, Kraft15-DAFX, He14-TASLPACM, Hashimoto18-AES, Avendano02-ICASSP, Allen77-JASA} in this same processing framework.
Some systems performed sub-optimally, while a group of methods including the proposed one performed in general equally well with only small per-item differences.
We expect this group of methods to perform comparably especially in application-oriented evaluation. 
Since we are more interested in showing that PAD-based up-mixing is ripe enough for applications rather than finding differences possibly stemming from unoptimized implementations, we include only the proposed method in the listening test.

\section{Results}
\label{sec:results}
\begin{figure}[tb]
\centering
\includegraphics[width=0.8\columnwidth]{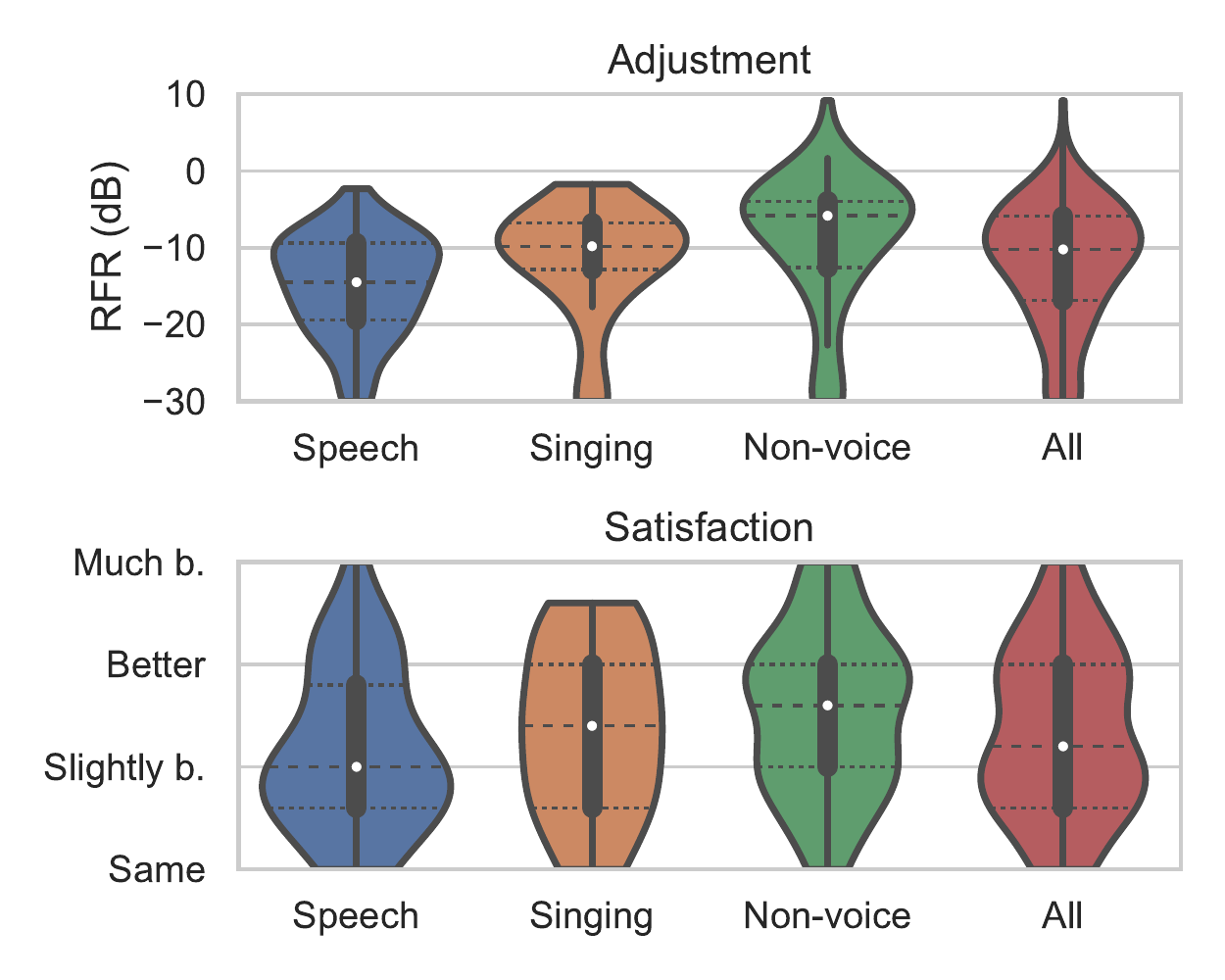}
\vspace{-0.5em}
\caption{Violin plots of the Rear-to-Front Ratios (RFR) selected during adjustment (initial RFR$=-\infty$\,dB truncated to -30\,dB) and resulting satisfaction levels for test items including speech, singing, not including any voice, and overall. 
The quartiles are shown by dashed lines. 
Medians for the three item groups are statistically significantly different both in terms of selected RFR and of resulting satisfaction. 
Overall, a median RFR of -10.2 dB is selected by the participants and the user satisfaction is clearly increased in every case (Wilcoxon signed rank test for median = \emph{Same}: $p < 0.00$).}
\label{fig:violin}
\vspace{-1em}
\end{figure}

The listening test involved 14 voluntary participants recruited among the research staff and interns at Fraunhofer IIS. 
The results of three of them were excluded in a post-screening phase, as they used the \emph{worse} part of the satisfaction scale at least once, indicating a misunderstanding of the given task or problems during the adjustment phase (the initial stereo mix could be selected during adjustment, so the final satisfaction should be at least \emph{same} as the initial condition). 
Figure~\ref{fig:violin} shows the main results after post-screening.
The adjustment is measured using Rear-to-Front Ratio (RFR), that is the energy ratio of the rear loudspeakers to the front loudspeakers, measured in dB:   
\begin{equation}
\mathrm{RFR} = 10 \log_{10} \frac{\sum_t \left(  | \remixLr(t) |^2 + | \remixRr (t) |^2 \right)}
{\sum_t \left( | \remixLf(t) |^2 + | \remixRf(t) |^2 \right)}.
\end{equation}
When analysing the listening test results, we observed three signal classes: \emph{speech} (7 items), \emph{singing} (3 items), and \emph{non-voice} (5 items) instead of the originally-assumed \emph{music} and \emph{broadcast}, and the result analysis uses this division.

It can be observed that the listeners made use of the possibility of re-locating some of the ambient component into the rear channels. 
Overall, a median RFR of -10.2\,dB is selected and the resulting satisfaction is clearly increased.   
The resulting up-mixed signals are rated by the participants on average being \emph{slightly better} to \emph{better} compared to the original stereo mix.
This suggests that the audio quality of the proposed approach is good enough for providing some level of up-mixing freedom that improves the satisfaction of the listeners compared to the original stereo audio.

Comparing items with speech, singing, or no voice at all, a trend appears. 
The highest RFRs (median  \mbox{-5.8\,dB}) are selected for non-voice items (instrumental music or applause), resulting in the highest satisfaction levels. 
Positive RFRs are also observed.
Speech items show the lowest RFRs (median \mbox{-14.5\,dB}). 
Music items with singing voice lie in between (median RFR \mbox{-9.8\,dB}).
Statistically significant difference between medians is tested by a Kruskal-Wallis test for both adjustment ($p < 0.00$) and satisfaction ($p = 0.026$), where non-parametric testing was selected because the Shapiro-Wilk test cast doubts on the normality of the considered groups.

The lower adjustment and satisfaction on speech items can be due to different factors. 
One explanation is that participants would desire higher RFRs also for these items, but had to compromise with decreasing quality due to inaccuracies introduced by the proposed algorithm (particularly well perceived on speech). 
An alternative is that higher RFRs are not aesthetically preferred on this kind of content. 
Reverberation associated with speech (or singing voice) can be a problem when it is moved as a part of the ambient component to the rear loudspeakers changing the perceived location of the talker (or the singer).
In fact, some participants reported that they did not want the talker position to change, while they enjoyed being surrounded by the scene background noise and music. 
Using a decorrelator on the signal moved in the rear channels could help on this issue, but this was omitted from the experiments for simplicity and reproducibility.

\section{Conclusion}
\label{sec:conclusions}
This paper presented a step-by-step derivation of a center-channel extraction method using MMSE-filters for the separation of the signal components. 
The method was extended into primary-ambient-decomposition with rotations for creating a centred scene with the primary component being the center channel, and the overall algorithm was simplified into two equations.
The proposed PAD method delivers results that were informally rated to be perceptually similar to solutions from the literature.
The method was evaluated using A/ST methodology in the application of stereo-to-quad up-mixing.
The test participants made use of the adjustment, especially for items without speech or singing, and their overall satisfaction to the spatial impression of the resulting up-mix signal was increased.

\bibliographystyle{IEEEtran}
\bibliography{PAD}

\end{document}